\documentclass[preprint,11pt]{elsarticle}

\usepackage[english]{babel}

\usepackage[letterpaper,top=2cm,bottom=2cm,left=3cm,right=3cm,marginparwidth=1.75cm]{geometry}

\usepackage{amsmath}
\usepackage{amssymb}
\usepackage{bm}
\usepackage{comment}
\usepackage{graphicx}
\usepackage{subcaption}
\usepackage[colorlinks=true, allcolors=blue]{hyperref}
\usepackage{adjustbox}
\usepackage{float}
\usepackage{tikz}
\usetikzlibrary{arrows.meta}
\usepackage{caption}
\usepackage{svg}
\usepackage{soul}
\usepackage{xcolor}
\usepackage{comment}
\usepackage{units}
\usepackage{ulem}
\usepackage{lineno}
\usepackage{tikz}
\usepackage{graphicx} 
\usepackage{subcaption}   
\usepackage{tcolorbox}
\usepackage{ulem}
\usepackage{cancel}
\usepackage{adjustbox} 
\usetikzlibrary{matrix} 
\newtcbox{\inlinecode}{on line, 
	colback=gray!10,  
	colframe=gray!50, 
	rounded corners,
	boxrule = 0.5pt,
	left = 1pt, right=1pt, top=0.5pt, bottom=0.5pt,
	fontupper = \ttfamily\small,
	boxsep=0pt
	}

\DeclareMathAccent{\svec}{\mathord}{letters}{126}

\journal{tbd}

\begin{document}

\begin{frontmatter}
\title{A novel parallelizable convergence accelerating method: Pointwise Frequency Damping}

\author[1]{Zikun Liu}
\author[2]{Xukun Wang}
\author[1]{Yilang Liu}
\author[1]{Weiwei Zhang}
\cortext[cor1]{Corresponding author}
\ead{aeroelastic@nwpu.edu.cn}

\address[1]{National Key Laboratory Science and Technology on Aerodynamic Design and Research, School of Aeronautics, Northwestern Polytechnical University, 710072, Xi'an, China}
\address[2]{ETSIAE-UPM-School of Aeronautics, Universidad Politécnica de Madrid, Plaza Cardenal Cisneros 3, E-28040 Madrid, Spain}


\begin{abstract}

This paper proposes a novel class of data-driven acceleration methods for steady-state flow field solvers. The core innovation lies in predicting and assigning the asymptotic limit value for each parameter during iterations based on its own historical data, rather than processing and assigning the entire flow field at once. This approach fundamentally guarantees identical results between serial and parallel computations. Subsequently, a formula for representing the asymptotic limit based on historical data is derived and discretized, yielding a purely algebraic expression.Furthermore, the applicability scope of the method is discussed, along with the underlying reasons for its acceleration capability. A quantitative expression for estimating the speedup ratio is also provided. Extensive validation cases were tested, ranging from the simplest inviscid airfoil flow to complex three-dimensional viscous transonic cruise flow around a aircraft, and solving asymmetric linear systems via GMRES. These tests consistently demonstrate significant acceleration effects with speedup factors ranging from 2.5 to 4. Combined with the near-zero computational overhead of the purely algebraic formulation during the solving process and the inherently parallel-compatible pointwise prediction principle, the results strongly indicate that this method is highly suitable for large-scale industrial mesh computations.

\end{abstract}

\end{frontmatter}

\tableofcontents
\section{Introduction}

Computational Fluid Dynamics (CFD) is a technique that employs numerical algorithms to solve the discretized governing equations of fluid mechanics. Over the past six decades, with advancements in computer hardware, this technology for studying fluid-related problems has rapidly gained widespread application and significantly advanced the field of fluid mechanics. In the 1960s, researchers started to solve potential flows governed by the Laplace equation for arbitrary geometries \cite{hess_calculation_1967}. In the 1970s, the need to solve transonic flows led to the development of methods to solve the small perturbation equation \cite{murman_calculation_1971}. With further progress in computing technology, the 1980s saw the feasibility of solving inviscid Navier-Stokes Equations(NSE), particularly with the emergence of the now widely used second-order accurate finite volume scheme \cite{jameson_numerical_1973}. Starting in the 1990s, CFD began addressing the Reynolds-Averaged Navier-Stokes Equations (RANS), which account for viscous effects in fluid flow. To this day, CFD applications largely remain at the RANS level, as progressing to Large Eddy Simulation (LES) or Direct Numerical Simulation (DNS) exceeds current computational capabilities \cite{witherden_future_2017}. Moreover, even for the steady-state RANS simulations in practical engineering applications, at least millions of degrees of freedom (DOFs) are typically required, posing a significant computational burden for existing hardware. To mitigate the limitation of computational resources, various classical convergence accelerating algorithms have been developed. 

To ensure numerical stability, implicit solution methods were developed, which introduced the challenge of solving large linear systems. Consequently, various efficient algorithms for solving linear equations were proposed, such as the Lower-Upper Symmetric Gauss-Seidel method (LU-SGS) \cite{jameson_lower-upper_1987}, the Generalized Minimal Residual method (GMRES) \cite{saad_gmres_1986}\cite{blanco_fast_1998}, and the Conjugate Gradient method (CG). Unlike using a uniform pseudo-time step globally, the local time-stepping method \cite{jameson_numerical_1981} selects the maximum allowable pseudo-time step based on local flow conditions in different regions, thereby accelerating convergence to a steady-state solution. In CFD solvers, high-frequency errors decay the fastest. The multigrid method exploits this property by solving the governing equations on grids of varying coarseness. Low-frequency errors that converge slowly on fine grids become high-frequency errors on coarse grids, accelerating their decay and improving overall convergence. This technique was first applied by Brandt to elliptic partial differential equations \cite{trottenberg_multigrid_2001} and later extended by Jameson to solve the Euler equations \cite{jameson_multigrid_1986}. Enthalpy damping \cite{jameson_numerical_1981} accelerates convergence by smoothing energy oscillations and decoupling the equation system, reducing numerical instability constraints on time-step size. This method is particularly effective for strongly nonlinear and multi-physics coupled problems. Jameson and Baker introduced residual smoothing \cite{enander_implicit_1995}, a technique designed to impart implicit characteristics to explicit schemes, significantly increasing the maximum allowable CFL number. Additionally, other acceleration techniques include low-Mach-number preconditioning \cite{WOS:A1987K481500001}, among others.

With the rise of artificial intelligence, data-driven acceleration methods have gradually become a research hotspot\cite{brunton_machine_nodate}. Unlike the classical algorithms mentioned above that directly improve solver efficiency, data-driven methods aim to utilize historical information generated during solver iterations to obtain more accurate solutions, thereby skipping some iteration steps to achieve acceleration. Data-driven acceleration is currently a relatively simple and easy-to-implement approach for CFD acceleration. It offers advantages such as easy code portability, relatively low computational requirements, and independence from grid structures and pseudo-time stepping schemes. One of the current research focuses in data-driven acceleration is using historical CFD iteration data from training datasets to train neural network (NN) models. During application, these trained NN models leverage partial historical data from CFD iterations to predict flow fields closer to steady-state solutions, thereby achieving acceleration effects\cite{zuo_fast_2024}\cite{hajgato_accelerating_2019}\cite{zhang_residual_2024}\cite{obiols-sales_cfdnet_2020}. However, this method generally requires that the cases in practical applications use the same mesh as the training data, meaning it cannot generalize to different geometries. Additionally, it relies on pre-collected CFD simulation cases for training and cannot be applied in real-time to new cases. We refer to this approach as offline data-driven acceleration. This method faces certain limitations: its effectiveness has been questioned, training neural networks demands specific hardware resources, and the training process is typically time-consuming. Currently, it still has some distance to go before practical application\cite{mcgreivy_weak_2024}.

In contrast to neural network-based data-driven methods, we refer to approaches that directly utilize historical data without training to obtain more accurate solutions as online data-driven acceleration methods. One of the earliest online data-driven techniques, proposed in the last century, is vector extrapolation\cite{ford_extrapolation_nodate}, including minimal polynomial extrapolation (MPE)\cite{cabay_polynomial_1976} and reduced-rank extrapolation (RRE)\cite{sidi_minimal_2017}. These methods approximate the converged solution through linear combinations of historical data sequences\cite{eddy_extrapolating_1979}\cite{mesina_convergence_1977}\cite{jbilou_vector_2000}. In the 1980s, vector extrapolation was successfully applied to CFD solvers, demonstrating effective acceleration\cite{duminil_fast_2014}\cite{hafez_applications_1987}\cite{sidi_convergence_1990}. Entering the 21st century, reduced-order models (ROMs) were developed for analyzing flow mechanisms, with the most well-known being proper orthogonal decomposition (POD)\cite{berkooz_proper_nodate} and dynamic mode decomposition (DMD)\cite{schmid_dynamic_2010}. Naturally, these flow-processing techniques were also adapted for accelerating CFD convergence\cite{leclainche_accelerating_2017}\cite{andersson_non-intrusive_2016}\cite{djeddi_convergence_2017}. For instance, Liu et al.\cite{liu_mode_2019} proposed the mode multigrid (MMG) method based on DMD, which filters high-frequency modes in pseudo-time and achieved acceleration in solving the Euler equations. This approach was later extended to accelerate RANS equations\cite{tan_improved_2021} and adjoint equations\cite{chen_accelerating_2020}. Additionally, Wang et al. introduced a mean-based minimal residual method, which employs a mean-based ROM to search for the flow field solution with minimal residuals in the residual space\cite{wang_data-driven_2024}.

In prior research, both offline and online data-driven acceleration methods uniformly processed the entire flow field before feeding the updated results back to the solver. While effective for small-scale CFD computations, this approach encounters significant challenges when applied to complex engineering problems requiring parallel computing. When the computational domain is divided into subdomains for parallel processing, separately processing each subdomain cannot guarantee identical results to full-field serial processing. This discrepancy compromises the accuracy of full-field methods and requires special treatment to maintain consistency at subdomain interfaces. Furthermore, matrix operations (such as DMD\cite{schmid_dynamic_2010}, POD\cite{berkooz_proper_nodate}, and SVD) typically have time complexities much greater than $O(n)$ for problems with $n$ variables. For matrices with tens of millions of elements, these methods become computationally intractable and may fail to converge for ill-conditioned matrices. In contrast, processing each grid point individually and then feeding the results back to the solver makes parallel computation particularly reasonable. Since we perform independent operations on each point, the serial and parallel results remain completely consistent without requiring consideration of information exchange between parallel subdomains.

The grid-point-based CFD acceleration approach follows the same fundamental concept as field-based processing methods - both process historical information to map to a flow field with reduced errors. The difference lies in that full-field processing seeks a matrix that reduces the norm of the residual matrix for the entire flow field, while the grid-point approach aims to find parameter values at each individual grid point that better approximate the steady-state solution. In other words, it involves finding the limiting values of the functions formed by each parameter's evolution with iteration steps at every grid point to achieve acceleration.

In this paper, we have mathematically derived an expression connecting initial values and limit values through infinite integration, which is rigorously valid in theory. However, since infinite integration cannot be practically implemented, we approximate the functional limit using truncated infinite integrals and discuss the applicable conditions of this formulation. When applying our derived formula to accelerate CFD computations, we recognized that the original formulation was designed for continuous functions. Therefore, we extended it to a time-discrete form, ultimately obtaining a purely algebraic expression through discretization. This algebraic nature ensures high computational efficiency in practical implementations. The discrete formula enables us to calculate parameter values at each grid point that better approximate the steady-state solution, thereby achieving acceleration. Notably, our implementation avoids inefficient loop structures for parameter iteration. Instead, by leveraging Python's element-wise vector operations (pointwise addition and multiplication), we can rapidly compute updated values for the entire flow field. Combined with the algebraic nature of our formulation, this approach achieves exceptional computational efficiency. For an iteration problem with $n$ variables, our algorithm maintains an $O(n)$ time complexity, with computational overhead approximately equivalent to just 1-2 pseudo-time iterations in standard CFD calculations. This efficiency makes our method particularly suitable for large-scale parallel computations while maintaining mathematical rigor.

The remainder of the paper is organised as follows. Section 2 details the proposed methodology. Subsequently, Section 3 discusses the applicability scope of the method and its acceleration principle. Section 4 then presents multiple numerical test cases, including CFD acceleration benchmarks and linear system solving acceleration experiments. Finally, Section 5 provides brief concluding remarks.

\section{Methodology}
\subsection{Formula derivation}\label{sec:formula_derivation}

Generally speaking, considering a dynamic system evolving to a steady-state, the variables on all the spacial positions will converge to some specific values. Therefore, the evolution of a variable on a fixed spacial position can be seen as a unary function $f$:
\begin{equation}
    f(t),\; t \in [0,+\infty),\; \vert \lim_{t\to +\infty}f(t)\vert<\infty.
\end{equation}

The basic idea of this paper is to fully use the historical data generated during the process of convergence. That is to say, our target is to establish a relation between $f(0^{+})$, $f(t)$ and $f(+\infty)$. First, we transfer $f(0^{+})$ and $f(+\infty)$ to integral form:
\begin{equation}
    f(0^{+}) = \lim_{\xi\to 0}f(\xi)=\frac{f(\xi)}{\ln{(b/a)}}\lim_{m\to 0}\int_{am}^{bm}\frac{1}{t}dt,\; \xi \in (am,bm),\; a,b \in \mathbb{Z^{+}}.
\end{equation}
Assuming $f \in C^0$ and applying mean value theorem inversely, it can be written as:
\begin{equation}
\label{eq:f_0+}
    f(0^{+}) = \frac{1}{\ln{(b/a)}}\lim_{m\to 0}\int_{am}^{bm}\frac{f(t)}{t}dt.
\end{equation}
Similarly, $f(+\infty)$ can also be written in integral form:
\begin{equation}
\label{eq:f_inf}
    f(+\infty)=\frac{f(\eta)}{\ln{(b/a)}}\lim_{M\to +\infty}\int_{am}^{bm}\frac{1}{t}dt = \frac{1}{\ln{(b/a)}}\lim_{M\to +\infty}\int_{aM}^{bM}\frac{f(t)}{t}dt,\; \eta \in (aM,bM),\; a,b \in \mathbb{Z^{+}}.
\end{equation}
Their difference can be obtained by subtracting Eq.\eqref{eq:f_inf} from Eq.\eqref{eq:f_0+}:
\begin{equation}
\label{eq:Delta1}
    \begin{split}
        \Delta &= f(0^+) - f(+\infty) \\
    & = \frac{1}{\ln{(b/a)}}\left( \lim_{m\to0}\int_{am}^{bm}\frac{f(t_1)}{t_1}dt_1 - \lim_{M\to \infty}\int_{aM}^{bM}\frac{f(t_2)}{t_2}dt_2 \right) \\
    & = \frac{1}{\ln{(b/a)}}\lim_{m\to0}\lim_{M\to \infty}\left(\int_{am}^{bm}\frac{f(t_1)}{t_1}dt_1 - \int_{aM}^{bM}\frac{f(t_2)}{t_2}dt_2\right).
    \end{split}
\end{equation}

Considering $\int_{bm}^{aM}f(t_1)/t_1dt_1 \equiv \int_{bm}^{aM}f(t_2)/t_2dt_2$, the upper and lower limit of integration in Eq.\eqref{eq:Delta1} can be exchanged:
\begin{equation}
\label{eq:Delta2}
\begin{split}
    \Delta &= \frac{1}{\ln{(b/a)}}\lim_{m\to0}\lim_{M\to \infty}\left( \int_{am}^{bm}\frac{f(t_1)}{t_1}dt_1 + \int_{bm}^{aM}f(t_1)/t_1dt_1 - \int_{bm}^{aM}f(t_2)/t_2dt_2 - \int_{aM}^{bM}\frac{f(t_2)}{t_2}dt_2 \right) \\
    & = \frac{1}{\ln{(b/a)}}\lim_{m\to0}\lim_{M\to \infty}\left( \int_{am}^{aM}\frac{f(t_1)}{t_1}dt_1 - \int_{bm}^{bM}\frac{f(t_2)}{t_2}dt_2\right).
\end{split}
\end{equation}
Substituting $t_1=at$ and $t_2=bt$ into Eq.\eqref{eq:Delta2}, the integration variable can be uniformed as:

\begin{equation}
\begin{split}
    \Delta &= \frac{1}{\ln{(b/a)}}\lim_{m\to0}\lim_{M\to \infty}\left( \int_{m}^{M}\frac{f(at)}{t}dt - \int_{m}^{M}\frac{f(bt)}{t}dt \right) \\
    & = \frac{1}{\ln{(b/a)}}\int_0^{+\infty}\frac{f(at)-f(bt)}{t}dt.
    \end{split}
\end{equation}
Finally, we get the expression of $f(+\infty)$:
\begin{equation}
\label{eq:formula1}
    f(+\infty) = f(0^+) - \frac{1}{\ln{(b/a)}}\int_0^{+\infty}\frac{f(a\tau)-f(b\tau)}{\tau}d\tau.
\end{equation}

\subsection{Spatio-temporal discrete system}

In numerical simulations, both spatial and temporal domain are discretised and the integration in infinite domain is impossible. In practical, we can only use the data up to a specific moment $t$ and truncate the original integration Eq.\eqref{eq:formula1} to get:
\begin{equation}
\label{eq:g}
    g(t) = f(0^+) - \frac{1}{\ln{(b/a)}}\int_0^t\frac{f(at)-f(bt)}{t}dt.
\end{equation}
where $g(t)$ is the predicted limit value using the data of original function $f(\tau),\tau\in [0,t]$.
The other problem of applying this formula is that only values on discrete moments can be used. Despite the assumption of continuity ($f\in C^0$) when deriving the formula, the truncated integration Eq.\eqref{eq:g} can be approximated by numerical quadrature using temporal-discretised values $f(t_i),i=0,1,2\dots.$

For time-discrete systems, to apply the aforementioned formula, one only needs to connect the corresponding parameter values between two adjacent pseudo-time steps using a linear polynomial. This constructs a continuous function, enabling the application of the formula.

Following this procedure, the constructed function can be expressed as:
\begin{equation}
f_i(x)=[f(i)-f(i-1)]x - f(i)i + f(i-1)i + f(i) \quad (x \in [i-1,i),, i=1,2,...)
\label{eq:f_discrete}
\end{equation}
where $f(i-1)$ and $f(i)$ represent the values of $f(x)$ at pseudo-time steps $i-1$ and $i$, respectively, generated by the iteration process. The formula involves two hyperparameters $a$ and $b$. We select $a=1$ and $b=2$ because integer values are computationally most convenient and this choice minimizes common multiples between $a$ and $b$, allowing fuller utilization of historical data.

The core expression is given by:
	\begin{equation}
		f(x)=f(0^+) - \lim _{m \rightarrow +\infty} \sum\limits_{i=0}\limits^{m} \frac{\int_i^{i+1} \frac{f_i(x)-f_i(2x)}{x} \, dx}{\ln 2}=f(0^+) - \lim _{m \rightarrow +\infty} \sum\limits_{i=0}\limits^{m} \frac{\int_i^{i+1} \frac{f_i(x)}{x}-\frac{f_i(2x)}{x} \, dx}{\ln 2}
		\label{eq:discrete_integration}
	\end{equation}

Simplification of $\int_i^{i+1} \frac{f_i(x)}{x} dx$:
	\begin{equation}
		\int_i^{i+1} \frac{f_i(x)}{x} = \sum_{i=0}^m \left( \int_i^{i+1} \left[ f(i\!+\!1) - f(i) \right] + \frac{f(i)(i\!+\!1) - f(i\!+\!1)}{\mathrm{x}} \, d\mathrm{x} \right)=In_{1}^{i}
		\label{eq:discretefx}
	\end{equation}

where:
\begin{equation}
		In_1^i = [f(i+1) - f(i)] + (f(i)(i+1) - f(i+1)i)\ln\left(1+\frac{1}{i}\right)
		\label{eq:discretefx_simplified}
	\end{equation}
Simplification of $\int_i^{i+1} \frac{f_i(2x)}{x} dx$:
\begin{equation}
		\begin{split}
		&\int_i^{i+1} \frac{f_i(2x)}{x}  = \int_i^{i+\frac{1}{2}} 2[f(2i+1) - f(2i)] + \frac{f(2i)(2i+1) - f(2i+1)2i}{x} \, dx \\
		&+ \int_{i+\frac{1}{2}}^{i+1} 2[f(2i+2) - f(2i+1)] - 	\frac{f(2i+1)(2i+2) - f(2i+2)(2i+1)}{x} \, dx  \\
        &= In_{2}^{i} + In_{2^{'}}^{i}
		\end{split}
	 \label{eq:discretef2x}
	\end{equation}
where:
	\begin{equation}
		\begin{split}
			In_{2}^{i} &= [f(2i+1) - f(2i)] + (f(2i)(2i+1) - f(2i+1)2i)\ln\left(1+\frac{1}{2i}\right)\\
			In_{2^{'}}^{i} &= [f(2i+2) - f(2i+1)] + [f(2i+1)(2i+2) - f(2i+2)(2i+1)]\ln\left(1+\frac{1}{2i+1}\right)
		\end{split}
		\label{eq:discretef2x_simplified}
	\end{equation}

Finite-Point Implementation:

In practical discrete systems, infinite points are infeasible. Given $2m+3$ discrete points, Eq.\eqref{eq:corn_discrete_finite} assumes the truncated form($a=1$, $b=2$):
	\begin{equation}
		f(+\infty) = f(0^+) - \lim _{m \rightarrow +\infty}\frac{\sum\limits_{i=0}^{m}{In_{1}^i} - {In_{2}^i} - {In_{2^{'}}^i}}{\ln 2}
		\label{eq:corn_discrete_finite}
	\end{equation}
	where the integral components are defined as:
	\begin{equation}
		\begin{split}
			In_1^i &= [f(i+1) - f(i)] + (f(i)(i+1) - f(i+1)i)\ln\left(1+\frac{1}{i}\right)\\
			In_2^i &= [f(2i+1) - f(2i)] + (f(2i)(2i+1) - f(2i+1)2i)\ln\left(1+\frac{1}{2i}\right)\\
			In_{2^{'}}^i &= [f(2i+2) - f(2i+1)] + [f(2i+1)(2i+2) - f(2i+2)(2i+1)]\ln\left(1+\frac{1}{2i+1}\right)
		\end{split}
		\label{eq:In}
	\end{equation}

\subsection{Application in CFD solver}
\label{sec:application_in_CFD}
	Following the preceding discussion, the functional limit can be approximated by Eq.\eqref{eq:g}. Applying this formula within a CFD solver can accelerate convergence. The procedure is as follows: Flow field snapshots are collected at intervals of $n_{skip}$ timesteps. All parameters within the $i$-th collected snapshot are concatenated into a column vector $V_i$. After gathering $2m+3$ snapshots, the following snapshot ensemble is formed:
	\begin{equation}
		\left\{
		\begin{matrix}
			 | &| & &|\\
			 V_{0}& V_{1},& \cdots ,&V_{2m+2} \\
			 | & | & &|
		\end{matrix}
		\right\}
		\label{eq:cfd_snaps}
	\end{equation}
	
	Eq.\eqref{eq:g}, originally defined for a single point, is reformulated in vector form to operate over the entire field, yielding Eq.\eqref{eq:corn_cfd}. Subsequently, Eq.\eqref{eq:corn_cfd} is applied once to the snapshot ensemble Eq.\eqref{eq:cfd_snaps} to generate a predicted flow field $V_{Pre}$, aiming to accelerate the iterative process.
	
	\begin{equation}
		V_{Pre} = V_0 - \frac{\sum\limits_{i=0}^{m}{In_{1}^i} - {In_{2}^i} - {In_{2^{'}}^i}}{\ln 2}
		\label{eq:corn_cfd}
	\end{equation}
	
	\begin{equation}
		\begin{split}
			In_1^i &= [V(i+1) - V(i)] + (V(i)(i+1) - V(i+1)i)\ln\left(1+\frac{1}{i}\right)\\
			In_2^i &= [V(2i+1) - V(2i)] + (V(2i)(2i+1) - V(2i+1)2i)\ln\left(1+\frac{1}{2i}\right)\\
			In_{2^{'}}^i &= [V(2i+2) - V(2i+1)] + [V(2i+1)(2i+2) - V(2i+2)(2i+1)]\ln\left(1+\frac{1}{2i+1}\right)
		\end{split}
		\label{eq:cfd_In}
	\end{equation}
	
	Since our approach processes each point in the flow field using exclusively its own historical data without incorporating information from other points, and employs a formula that utilizes historical data to forecast the functional limit, we designate this category of methods under the general term: Pointwise Limit-Forecasting (PLF).

\section{Discussion on scope of application}

\subsection{Additional conditions for practical accelerating}

Although the integration Eq.\eqref{eq:formula1} is valid for all continuous functions defined on $[0,+\infty)$ existing one-side limit, considering that only finite data from $f(t)$ can be used, more requirements have to be imposed on $f$ to make sure that the transformed function $g(t)$ is closer to $f(+\infty)$ than original $f(t)$ for finite $t$. Here, we consider functions with the following specific form:
\begin{equation}
\label{eq:specificfunc}
    f(t) = \frac{a}{x^m}\sin(nx+m) + b
\end{equation}
where $a, b, m$ and $n$ are all finite real numbers. By taking $a=1$ and $b=2$ in Eq.\eqref{eq:g}, we have:
\begin{equation}
    g(t) = f(0^+) - \frac{1}{\ln2}\int_0^t\frac{f(\tau) - f(2\tau)}{\tau}d\tau.
\end{equation}
Assuming $ df/dt \neq 0$, we can define the following ratio:
\begin{equation}
\label{eq:ratio}
    R(t) = \frac{\vert dg/dt\vert }{\vert df/dt \vert} = \frac{\vert f(t) - f(2t)\vert}{\vert tdf(2t)/dt\vert}.
\end{equation}
If $\lim_{t\to+\infty}R(t)=0$, we can come to the conclusion that the convergence order of transformed function $g(t)$ is higher than that of original function $f(t)$, which means $g(t)$ convergences to the limit faster than $f(t)$.
Substituting Eq.\eqref{eq:specificfunc} into Eq.\eqref{eq:ratio}, we obtain the following:
\begin{equation}
    R(t) = \frac{2^{m+1}\sin(nx+m)-2\sin(2nx+m)}{2nx\cos(2nx+m)-m\sin(2nx+m)},
\end{equation}
where $R(t)$ converges to $0$ as $t$ intends to infinity, except at points $X=\{x\;\vert\; x=(\pi/2+k\pi)/{2n}-m,\;k\in \mathbb{Z}\}$. However, considering the Lebesgue measure of this set $\mu(X)=0$, we can get the conclusion that $R(t)$ converges to zero on almost all the points when $t$ tends to infinity, which means $g(t)$ has higher convergence order than $f(t)$.

\subsection{Effects on spatio-temporal system}
\label{sec:effects_on_spatio-temporal_system}
In the previous subsection, we discuss the effectiveness of proposed method in case of unary function (neglecting the spatial dimension effects). However, in most practical computational physics scenario, solutions are defined in both temporal and spatial domain. In this section, the effects of PFD on spatio-temporal system will be analyzed. Before the analysis, we have to simplify the formula of $g(t)$ in Eq.\eqref{eq:g}. Noting that the numerator in the integral can be written as:
\begin{equation}
    f(at) - f(bt) = -\int_a^b\frac{df(st)}{d(st)}tds,
\end{equation}
and the expression of $g(t)$ can be modified by exchanging the integration order:
\begin{equation}
\begin{split}
     g(t) &= f(0^{+})+ \frac{1}{\ln{(b/a)}}\int_0^t\int_a^b\frac{df(s\tau)}{d(s\tau)}dsd\tau \\
     \label{eq:c0}
     &= f(0^{+})+ \frac{1}{\ln{(b/a)}}\int_a^b\frac{1}{s}\int_0^{st}\frac{df(s\tau)}{d(s\tau)}d(s\tau)ds
\end{split}
\end{equation}
where the integral inside can be written as:
\begin{equation}
\label{eq:c1}
    \int_0^{st}\frac{df(s\tau)}{d(s\tau)}d(s\tau) = f(st) - f(0)
\end{equation}
and the entire integral reduces to:
\begin{equation}
\label{eq:c2}
    \int_a^b\frac{f(st) - f(0)}{s}ds = \int_a^b\frac{f(st)}{s}ds - \ln(b/a)f(0).
\end{equation}
Combining Eq.\eqref{eq:c0}, Eq.\eqref{eq:c1} and Eq.\eqref{eq:c2} and setting $u=st$, $g(t)$ can be further simplified as:
\begin{equation}
\label{eq:g2}
    g(t) = \frac{1}{\ln{(b/a)}}\int_{at}^{bt}\frac{f(u)}{u}du.
\end{equation}

Now we consider the evolution of the one-dimensional error $Er(x,t):\mathbb{R}\times\mathbb{R}^{+}\to\mathbb{R}$ and assume that it can be expanded as follows:
\begin{equation}
\label{eq:infiniteEr}
    Er(x,t)=\sum_{i=0}^{+\infty}e^{-\sigma_it}\cos(\omega_it+k_ix).
\end{equation}
We truncate the high-order components ($i>m$) in the infinite series Eq.\eqref{eq:infiniteEr} to get: 
\begin{equation}
\label{eq:Er}
    Er(x,t) \approx \sum_{i=0}^m Er_i(x,t)
\end{equation}
where $Er_i(x,t) = e^{-\sigma_it}\cos(\omega_it+k_ix)$, $\sigma_i$ is the decay rate, $\omega_i$ and $k_i$ denote the frequency in the temporal and spatial domain, respectively. 

The method we proposed is termed as PFD and the reason is detailed in the end of this subsection. Substituting Eq.\eqref{eq:Er} into Eq.\eqref{eq:g2}, we can get the expression of transformed error, $Er^{\text{PFD}}(x,t)$. Without loss of generality, we set $a=0.5$ and $b=1$ to keep the same ratio between $a$ and $b$ as before:
\begin{equation}
\label{eq:ErPFD1}
    Er^{\text{PFD}}_i(x,t) = \frac{1}{\ln2}\int_{t/2}^t\frac{e^{-\sigma_i\tau}\cos(\omega_i\tau+k_ix)}{\tau}d\tau
    = \frac{1}{\ln2}\Re \left[e^{\imath k_i x}\int_{t/2}^t\frac{e^{(-\sigma_i+\imath\omega_i)\tau}}{\tau}d\tau \right].
\end{equation}
Introducing the exponentially integral $\text{Ei}$:
\begin{equation*}
    \text{Ei}(z) =\int_{-\infty}^z\frac{e^\tau}{\tau}d\tau, \; z\in \mathbb{C},
\end{equation*}
the integral in Eq.\eqref{eq:ErPFD1} becomes:
\begin{equation}
\label{eq:C}
    \int_{t/2}^t\frac{e^{(-\sigma_i+\imath\omega_i)\tau}}{\tau}d\tau = \text{Ei}\left( (-\sigma_i+\imath\omega_i)t \right) - \text{Ei}\left( (-\sigma_i+\imath\omega_i)t/2 \right) = C.
\end{equation}
Here we denote this integral as $C=A+\imath B$ where $A$ and $B$ are the real and imaginary part of $C$ respectively. Therefore, the transformed error can be written as:
\begin{equation}
    Er^{\text{PFD}}_i(x,t) = \frac{1}{\ln2}\Re\left[ e^{\imath k_ix}(A+\imath B) \right] = \frac{1}{\ln2}\left[ A\cos(k_i x) - B\sin(k_ix)\right].
\end{equation}
Assuming that for the error at moment $t_0$, all the components $Er_i(x,t_0)$ are supported on the compact set $[-\frac{2\pi}{\omega},\frac{2\pi}{\omega}] $ and $L_2$-integrable on this set, we can compare the $L_2$-norm of original error with that of the transformed error at moment $t_0$.

According to the formula of the original error $Er(x,t)$, its $L_2$-norm at $t=t_0$ is:
\begin{equation}
    \vert Er_i(x,t_0)\vert_2 = \sqrt{\int_{-2\pi/\omega_i}^{2\pi/\omega_i}\left( e^{-\sigma_i t_0}\cos(\omega_it_0+k_ix) \right)^2dx} = e^{-\sigma_it_0}\sqrt{\frac{2\pi}{k_i}}.
\end{equation}
Similarly, the $L_2$-norm of transformed error at $t=t_0$ is:
\begin{equation}
	\begin{split}
		&\vert Er_i^{\text{PFD}}(x, t_0)\vert_2 = \sqrt{\int_{-2\pi / \omega_i}^{2\pi / \omega_i} \left[ \frac{1}{\ln 2} (A \cos(k_i x) - B \sin(k_i x)) \right]^2  dx} \\
		&= \frac{1}{\ln 2} \sqrt{ \int_{-2\pi / \omega_i}^{2\pi / \omega_i} A^2 \cos^2(k_i x)  dx + \int_{-2\pi / \omega_i}^{2\pi / \omega_i} B^2 \sin^2(k_i x)  dx - 2AB \int_{-2\pi / \omega_i}^{2\pi / \omega_i} \cos(k_i x) \sin(k_i x)  dx } \\
		&= \frac{1}{\ln 2} |C| \sqrt{\frac{2\pi}{k_i}}.
		\end{split}
\end{equation}
We define a scaling factor $\delta$ by the ratio between these two $L_2$-norm:
\begin{equation}
\label{eq:delta}
    \delta = \frac{\vert Er_i^{\text{PFD}}(x, t_0)\vert_2}{\vert Er_i(x,t_0)\vert_2} = \frac{\vert C \vert}{e^{-\sigma_it_0}\ln2}
\end{equation}
and the iteration is considered as being accelerated if $\delta <1$.

Recalling that $C$ is defined by $\text{Ei}$, to compute $\delta$ explicitly, $\text{Ei}$ can be expanded using asymptotic analysis:
\begin{equation}
	\operatorname{Ei}(z) = -\frac{e^z}{z} \left( 1 - \frac{1}{z} + \cdots + \frac{n!}{(-z)^n} + \cdots \right) = -\frac{e^z}{z} \sum_{n=0}^{\infty} \frac{n!}{(-z)^n},
	\label{eq:Asymptotic Analysis}
\end{equation}
and for the case $\vert z \vert \gg 1$, it can be approximated by its first principal term:
\begin{equation}
	\operatorname{Ei}(z) \approx -\frac{e^z}{z}.
	\label{eq:approx_Asymptotic Analysis}
\end{equation}
Substituting Eq.\eqref{eq:approx_Asymptotic Analysis} into Eq.\eqref{eq:C}, we have:
\begin{equation}
	C = \operatorname{Ei}(z_1) - \operatorname{Ei}(z_2) \approx -\frac{e^{z_1}}{z_1} + \frac{e^{z_2}}{z_2},
\end{equation}
where $z_1 = (-\sigma_i+\imath\omega_i)t$ and $z_2 = (-\sigma_i+\imath\omega_i)t/2$ and $\vert C \vert$ can be written as:
\begin{equation}
    \vert C \vert = \frac{e^{-\sigma_i t_0}}{t_0 \sqrt{\sigma_i^2 + \omega_i^2}} \left| e^{\imath \omega_i t_0} - 2e^{-\sigma_i t_0/2+\imath \omega_i t_0/2} \right|.
	\label{eq:C_mode}
\end{equation}
Combining Eq.\eqref{eq:C_mode} and Eq.\eqref{eq:delta}, we will get:
\begin{equation}
	\delta = \frac{ \left| e^{-\imath \omega_i t_0} - 2e^{\frac{(\sigma_i - \imath \omega_i) t_0}{2}} \right| }{t_0 \sqrt{\sigma_i^2 + \omega_i^2} \ln 2}.
	\label{eq:delta_scale_simplified}
\end{equation}
The scaling factor can be amplified using Cauchy-Schwarz inequality:
\begin{equation}
	\delta \leq \frac{ \left| e^{-i\omega_i t_0} \right| + 2 \left| e^{-\frac{(\sigma_i - i\omega_i) t_0}{2}} \right| }{t_0 \sqrt{\sigma_i^2 + \omega_i^2} \ln 2} = \frac{1 + 2e^{-\frac{\sigma_i t_0}{2}} }{t_0 \sqrt{\sigma_i^2 + \omega_i^2} \ln 2}. 
\end{equation}
Finally, the error is reduced if the following condition is satisfied:
\begin{equation}
	\Delta (\sigma_i, \omega_i, t_0) = \frac{1 + 2e^{\frac{\sigma_i t_0}{2}} }{t_0 \sqrt{\sigma_i^2 + \omega_i^2} \ln 2}<1.
	\label{eq:Delta}
\end{equation}

As for the discriminant Eq.\eqref{eq:Delta}, two conclusions can be get:
\begin{enumerate}
    \item Fixing $\sigma_i$  and $t_0$ (decay rate and sampling window), $\Delta (\sigma_i, \omega_i, t_0)$ tends to zero as the frequency $\omega_i$ increases, which means that the PFD method is more effective to reduce high-frequency error.
    \item  Fixing $\omega_i$ and $t_0$ (frequency and sampling window), $\Delta (\sigma_i, \omega_i, t_0) \ll 1$ when the decay rate $\sigma_i$ tends to $0$. In addition, the following statement can be proved considering the continuity:
    \begin{equation}
        \exists \sigma^{\ast}>0, \forall \sigma_i\in (0,\sigma^{\ast}),\; s.t.\; \Delta (\sigma_i, \omega_i, t_0)<1.
    \end{equation}
\end{enumerate}
In summary, the PFD method is effective in reducing the error of high frequency and low decay rate. 

\begin{figure}[h]
    \centering

    \begin{subfigure}[b]{0.49\textwidth}  
        \centering
        \includegraphics[width=\textwidth]{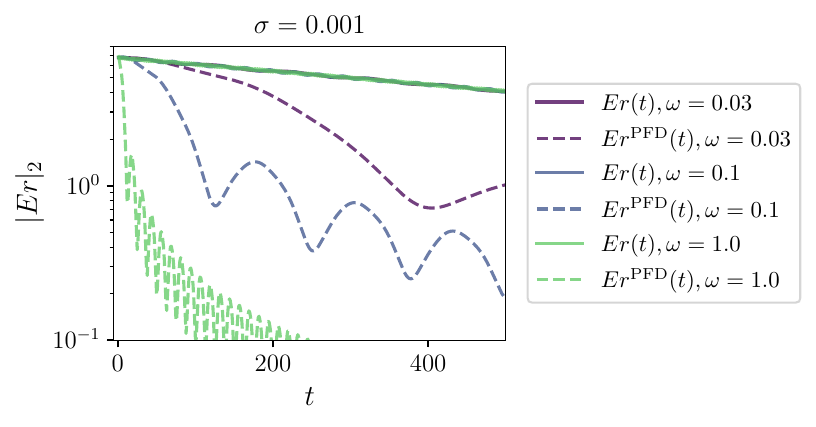}  
        \caption{The effects of different $\omega$ for fixed $\sigma$.}
        \label{fig:sigma=0.001}
    \end{subfigure}
    \hfill  
    \begin{subfigure}[b]{0.49\textwidth}
        \centering
        \includegraphics[width=\textwidth]{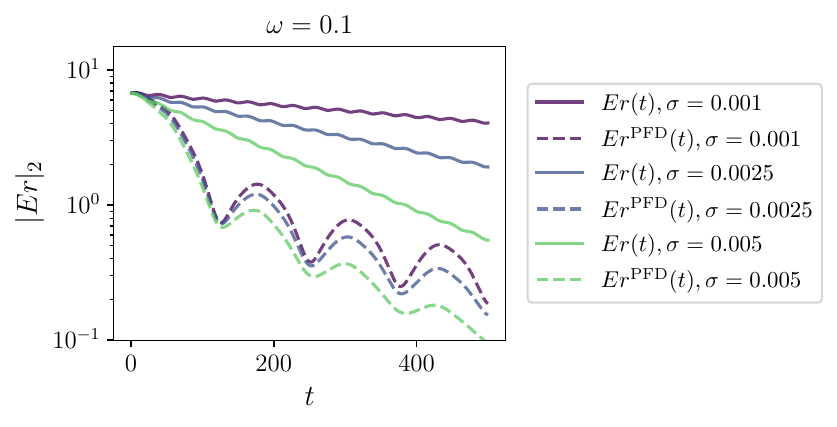}  
        \caption{The effects of different $\sigma$ for fixed $\omega$.}
        \label{fig:omega=0.1}
    \end{subfigure}

    \caption{Comparison of evolutions of $Er$ and $Er^{\text{PFD}}$ for different $\sigma$ and $\omega$.}
    \label{fig:comparison_evolutions_L2}
\end{figure}

The following is a simple numerical validation test.Considering a model single-frequency error evolution:
\begin{equation}
    Er(x,t) = e^{-\sigma t} \cos(\omega t+x),
\end{equation}
we apply the transform for different $\sigma$ and $\omega$ to validate our conclusions. First, we fix the decay rate $\sigma=0.001$ and test the acceleration effects of PFD for $\omega \in [0.03, 0.1, 1.0]$. The comparison of evolutions of error are plotted in Fig.\ref{fig:sigma=0.001}, where it can be seen clearly that the PFD method has better accelerating effects for high-frequency error components. Next, we study the effects of different decay rates $\sigma$ for the same frequency $\omega$ in Fig.\ref{fig:omega=0.1}, where $\omega=0.1$ and $\sigma\in[0.001, 0.0025, 0.005]$. As we expected, the smaller the decay rate, the larger the ratio of accelerating of PFD.

\begin{figure}[h]
	\centering
	\includegraphics[width=0.6\linewidth]{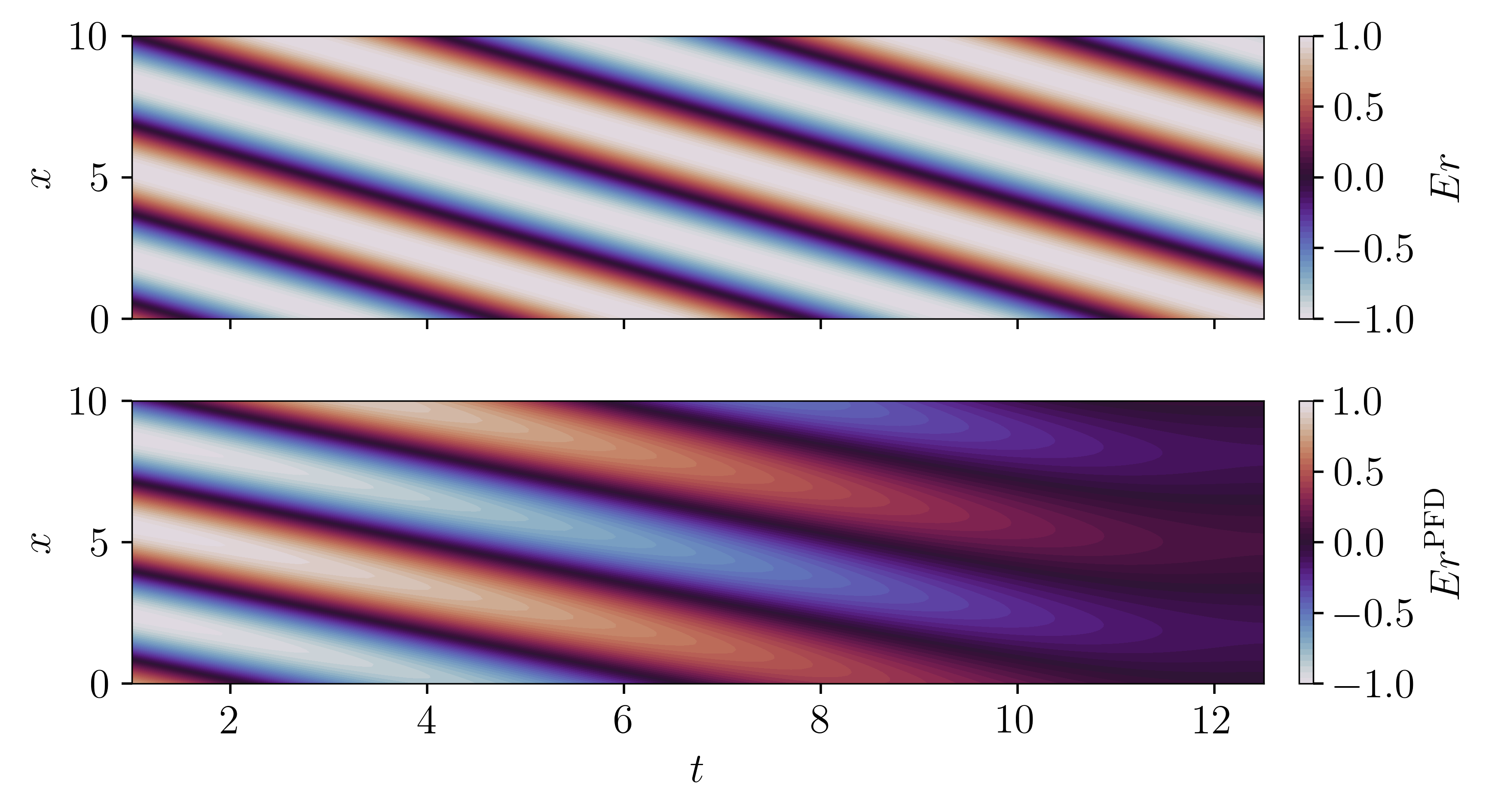} 
	\caption{Comparison of the contours of $Er$ and $Er^{\text{PFD}}$ for $\sigma=0.001$, $\omega = 1$ and $k=1$.}
	\label{fig:comparison_evolutions_contours}
\end{figure}

In addition, to show the damping effects of PFD in spatio-temporal domain, the $t-x$ contours of original error $Er$ and transformed error $Er^{\text{PFD}}$ for $\sigma=0.001$ and $\omega=1.0$ are plotted in Fig.\ref{fig:comparison_evolutions_contours}. It can be seen that the amplitude of oscillation is greatly damped by PFD, illustrating the obvious accelerating effects.

From the derivation and the simple numerical test, it can be found that the proposed method is very effective on damping the oscillating errors of high-frequency. Consider the pointwise feature of this method, we term is as Pointwise Frequency Damping, PFD in short.

\section{Results}\label{sec:results}

\subsection{Case 1: Subsonic inviscid flow over NACA0012 airfoil}
\label{sec:Euler}
During the CFD computation process, if the evolutionary history of variables at grid points (generated over iterations) can be decomposed to contain "modes" resembling functions like \eqref{eq:specificfunc}, then a single application of the PFD prediction will yield a value substantially closer to the value corresponding to a converged CFD solution. This constitutes an "improved parameter estimate." When the majority of parameter values achieve such "improved estimates," the residual of the flow field should consequently decrease significantly, thereby achieving the effect of accelerated convergence.

\begin{figure}[h]
    \centering

    \begin{subfigure}[b]{0.48\textwidth}  
        \centering
        \includegraphics[width=\textwidth]{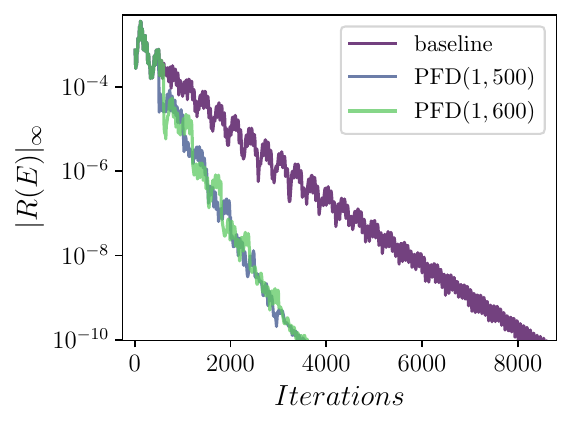}  
        \caption{Convergence histories of residual}
        \label{fig:resi_euler}
    \end{subfigure}
    \hfill  
    \begin{subfigure}[b]{0.49\textwidth}
        \centering
        \includegraphics[width=\textwidth]{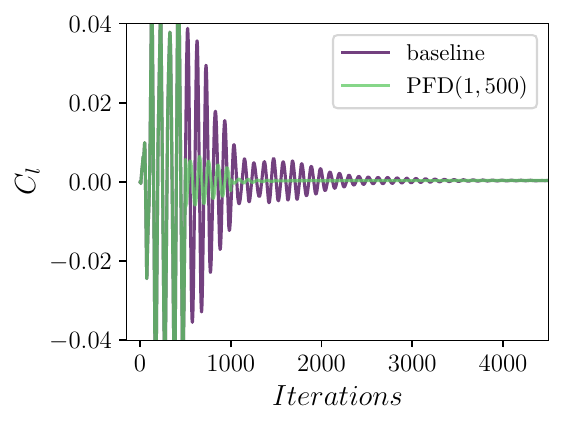}  
        \caption{Convergence histories of $C_l$}
        \label{fig:Cl_euler}
    \end{subfigure}

    \caption{Comparison of convergence histories of baseline method and PFD-accelerated methods for the case of subsonic inviscid flow over NACA0012 airfoil.}
    \label{fig:mesh_pressure_euler}
\end{figure}

In this test case, the convergence criterion is set as the residual dropping to $10^{-10}$. The figure presents the pressure contour of the converged flow field. Within this paper, the results computed by the original CFD solver serve as the benchmark method, and the resulting convergence curve is termed the baseline. Fig.\ref{fig:resi_euler} compares the PFD method with this benchmark method. In the legend, the two parameters following "PFD" denote the snapshot sampling interval and the total number of snapshots, respectively. Their product defines the window size for one application of the PFD method. For instance, PFD(20,50) indicates that one flow field snapshot is sampled every 20 pseudo-time steps, and the PFD method is applied once after collecting 50 snapshots.

Fig.\ref{fig:resi_euler} compares the residual convergence curves. Here, two PFD configurations with total snapshot counts of 400 and 500, respectively, are compared against the original convergence curve. Both configurations achieve similar acceleration effects, reducing the number of iteration steps by 55\% compared to the original iterative method. According to the definition of the acceleration ratio in Section \ref{sec:acc_rate_scale}, the original method requires 8250 iterations for the residual to reach $10^{-10}$, while the PFD-augmented method requires only 3750 iterations, producing an acceleration ratio of 2.20.

Fig.\ref{fig:Cl_euler} compares the lift coefficient per unit span ($C_L$) for the NACA0012 airfoil ($C_L = \frac{L}{\frac{1}{2} \rho V^2 b}$, where $L$ is the lift force, $\rho$ is the air density, $V$ is the freestream velocity, and $b$ is the reference length; for unit span, $b$ = 1). After incorporating the PFD method, the lift coefficient exhibits no significant oscillations beyond 2500 iterations. In contrast, the original convergence method still shows noticeable oscillations in the lift coefficient at 6000 iterations. The convergence curve of the lift coefficient demonstrates that the PFD method effectively suppresses oscillations during convergence, enabling the lift coefficient to reach its converged flow field value more rapidly. This observation further corroborates the acceleration capability of PFD for steady flow field computations.

\begin{figure}[h]
    \centering

    \begin{subfigure}[b]{0.49\textwidth}  
        \centering
        \includegraphics[width=\textwidth]{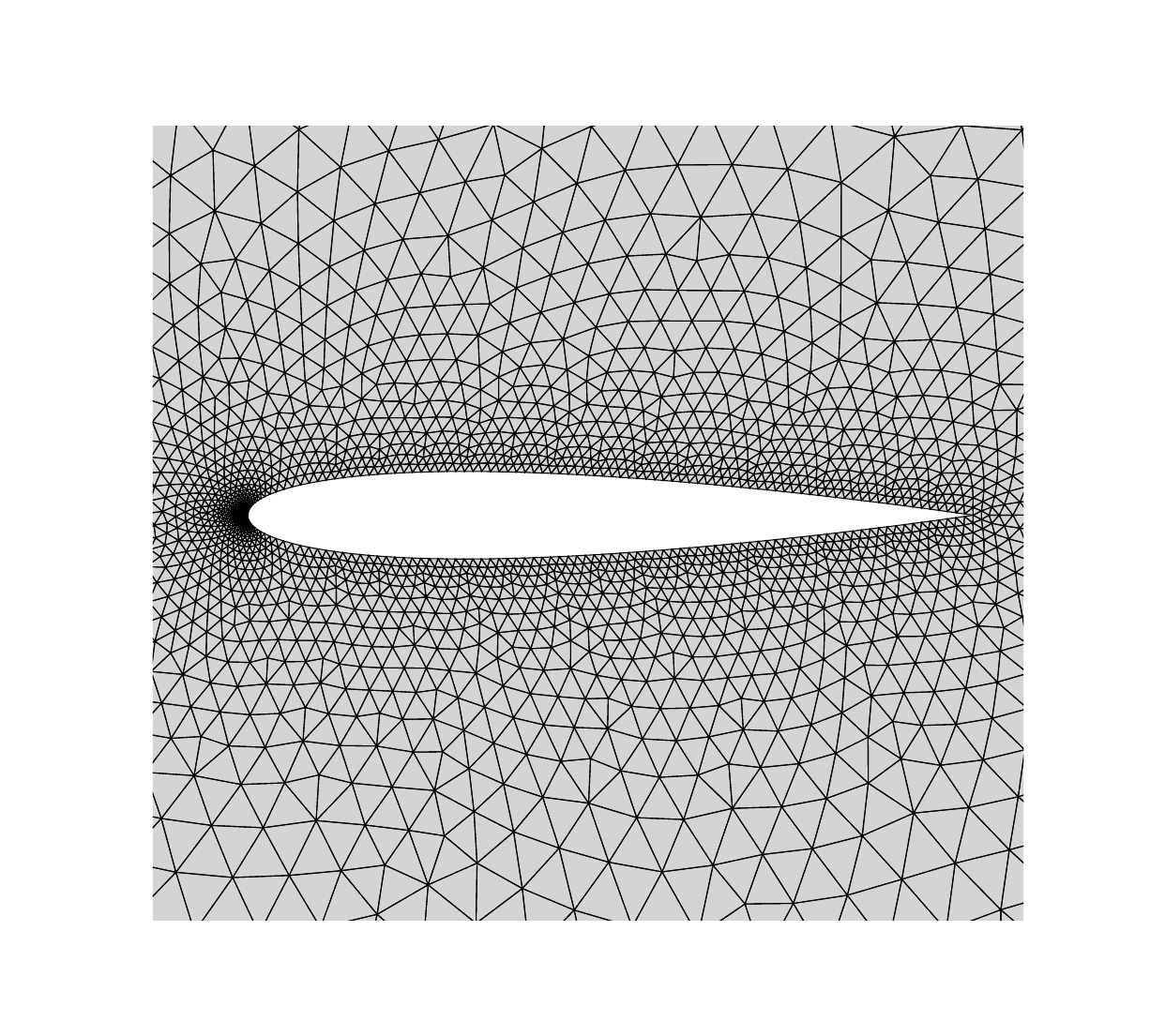}  
        \caption{Computational grid}
        \label{fig:mesh_euler}
    \end{subfigure}
    \hfill  
    \begin{subfigure}[b]{0.49\textwidth}
        \centering
        \includegraphics[width=\textwidth]{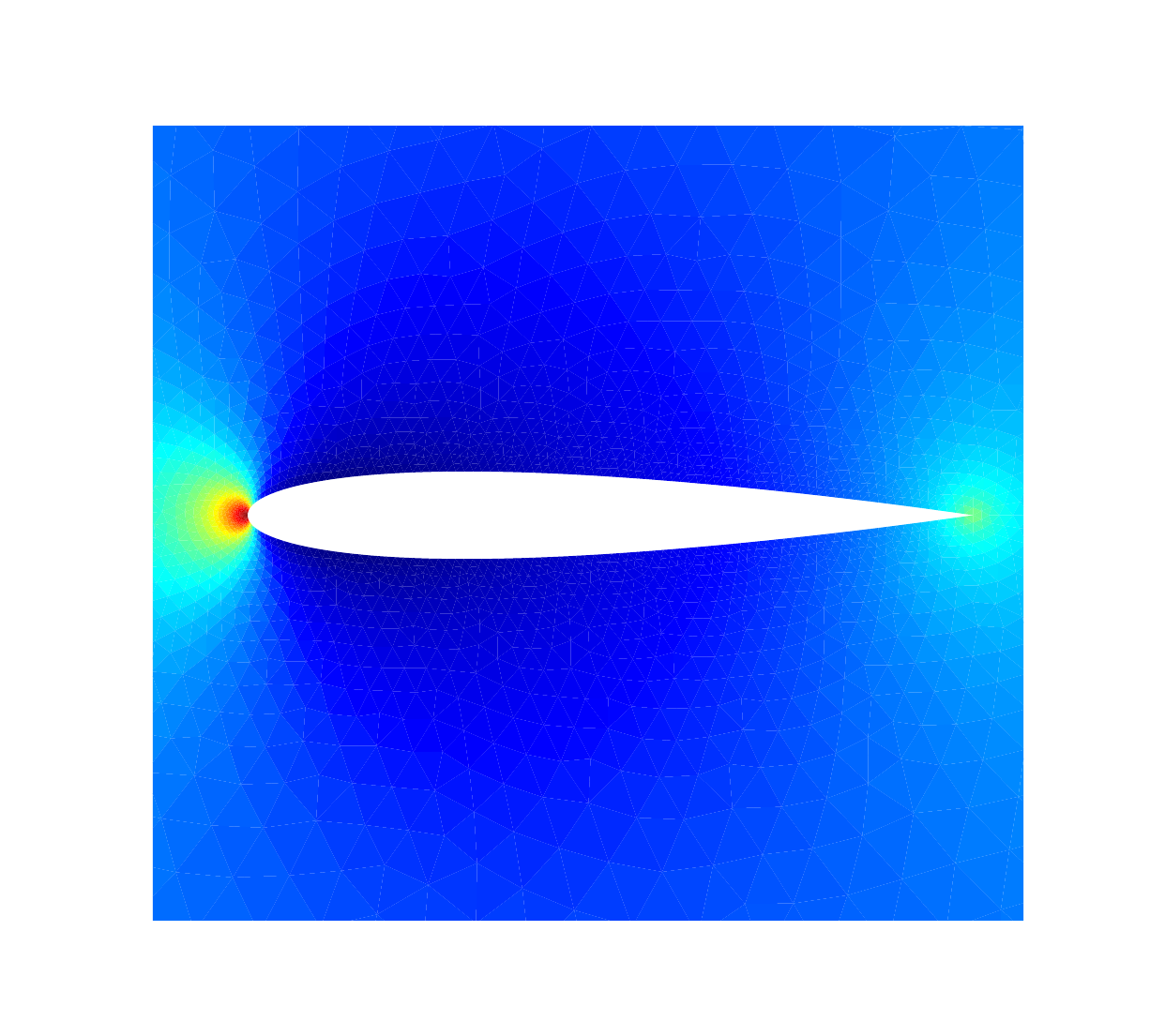}  
        \caption{Pressure distribution}
        \label{fig:flowfiled_euler}
    \end{subfigure}

    \caption{Computational grid and pressure distribution of inviscid flow over NACA0012 airfoil.}
    \label{fig:convergence_histories_euler}
\end{figure}

Synthesizing the results from Fig.\ref{fig:convergence_histories_euler}, it can be concluded that the acceleration capability of PFD manifests in reducing the residual of the computed flow field and obtaining converged force coefficients more rapidly.

\subsection{Case 2: Subsonic laminar flow over NACA0012 airfoil}

\begin{figure}[h]
    \centering

    \begin{subfigure}[b]{0.49\textwidth}  
        \centering
        \includegraphics[width=\textwidth]{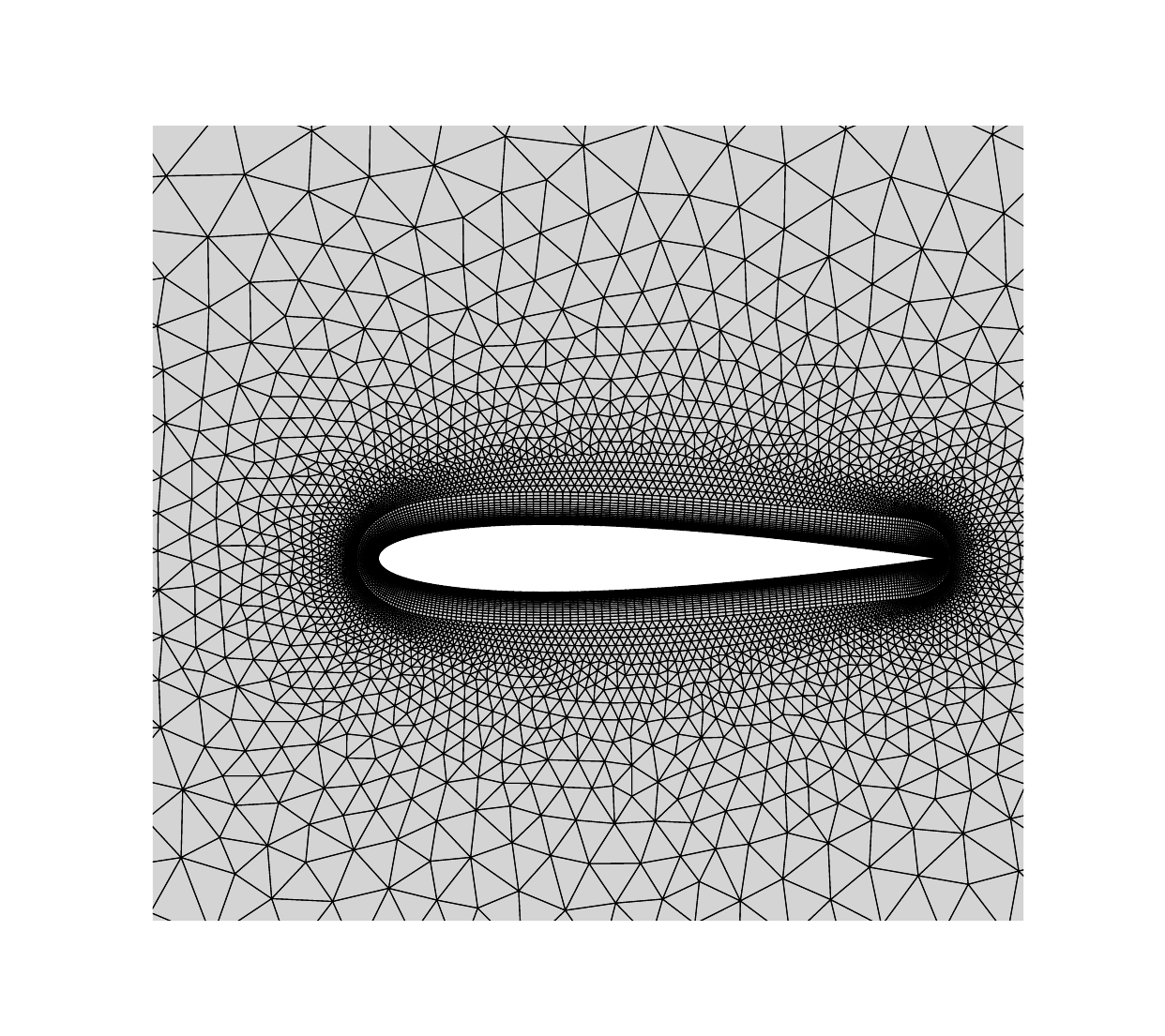}  
        \caption{Computational grid}
        \label{fig:mesh_laminar}
    \end{subfigure}
    \hfill  
    \begin{subfigure}[b]{0.49\textwidth}
        \centering
        \includegraphics[width=\textwidth]{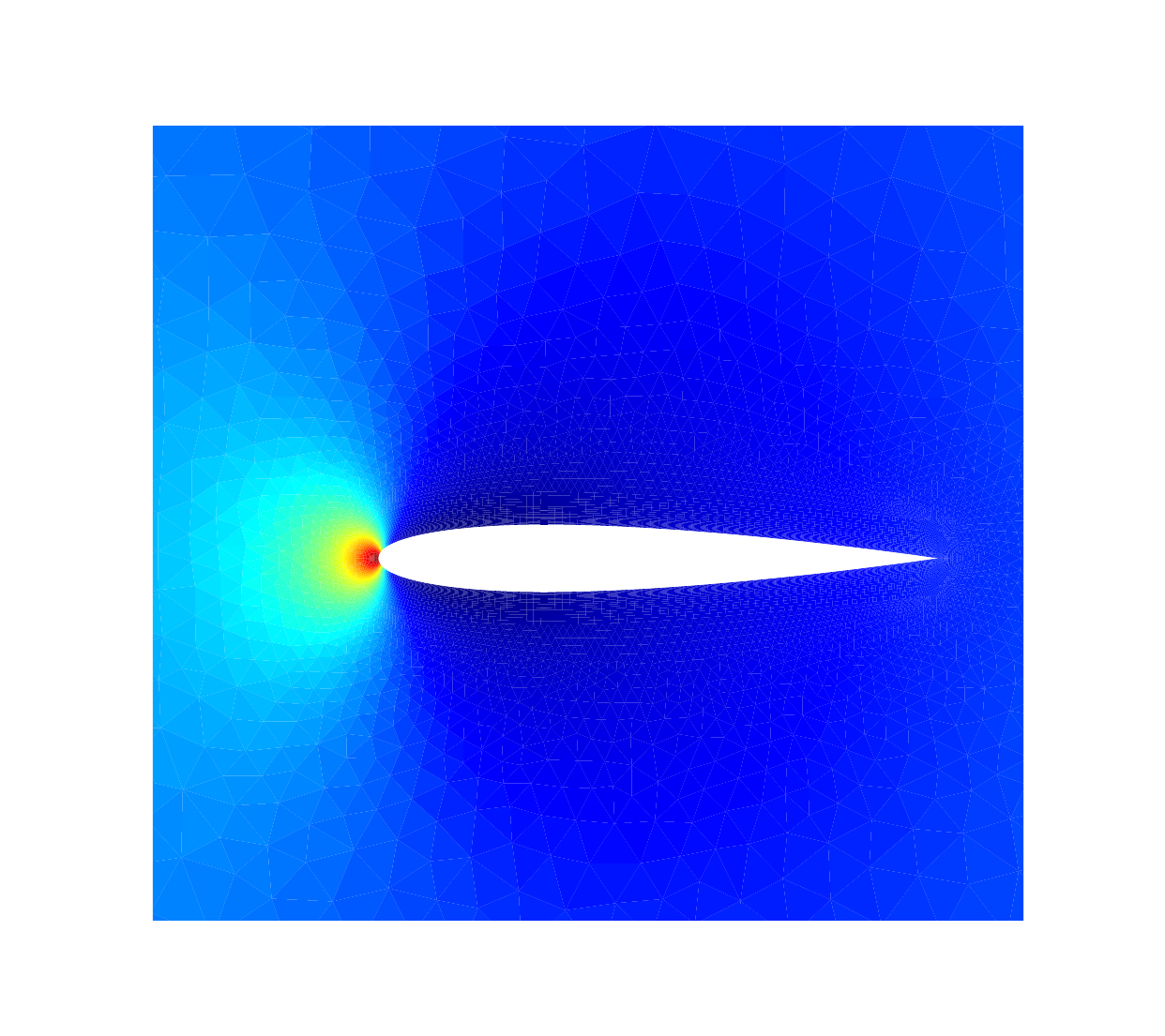}  
        \caption{Pressure distribution}
        \label{fig:flowfiled_laminar}
    \end{subfigure}

    \caption{Computational grid and pressure distribution of laminar flow over NACA0012 airfoil.}
    \label{fig:mesh_pressure_laminar}
\end{figure}

In Section \ref{sec:Euler}, we presented the inviscid $NACA0012$ test case. Below, we test the effectiveness of the PFD method for a more complex laminar flow. In this case, the computational grid, shown in Fig.\ref{fig:mesh_laminar}, consists of $15,675$ cells, is unstructured, and employs a CFL number of $2$. The freestream conditions are a Mach number of $0.5$, a Reynolds number of $5000$, and an airfoil angle of attack of $0°$. The pseudo-time integration scheme is the implicit Symmetric Gauss-Seidel (SGS) scheme.

\begin{figure}[h]
    \centering

    \begin{subfigure}[b]{0.48\textwidth}  
        \centering
        \includegraphics[width=\textwidth]{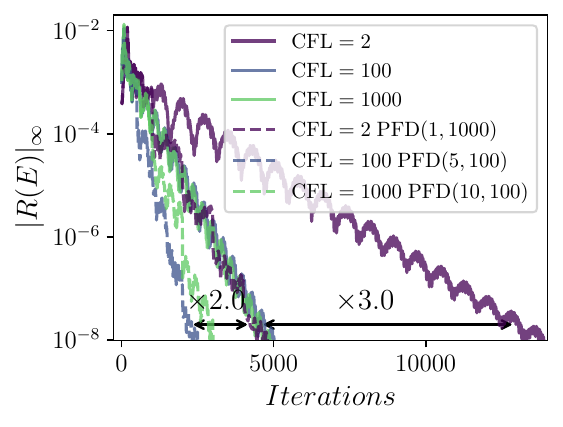}  
        \caption{Convergence histories of residual}
        \label{fig:resi_laminar}
    \end{subfigure}
    \hfill  
    \begin{subfigure}[b]{0.50\textwidth}
        \centering
        \includegraphics[width=\textwidth]{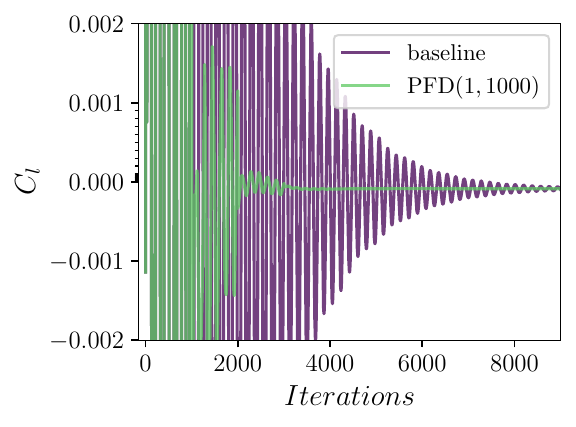}  
        \caption{Convergence histories of $C_l$}
        \label{fig:Cl_laminar}
    \end{subfigure}

    \caption{Comparison of convergence histories of baseline method and PFD-accelerated methods for the case of subsonic laminar flow over NACA0012 airfoil.}
    \label{fig:convergence_histories_laminar}
\end{figure}

The convergence criterion remains the residual dropping to $10^{-10}$. The original solver requires $20,000$ iterations to meet this criterion, as shown in Fig.\ref{fig:resi_laminar}. Here, we compare the PFD method configured as PFD(1,1000) against the original solver. It can be observed that incorporating the PFD method significantly reduces the number of iterations required to reach the converged state. Convergence is achieved in only $6,500$ iterations, representing a reduction of $67.5\%$. Using the acceleration ratio definition provided in Section \ref{sec:Euler}, the acceleration ratio is $3.08$. Fig.\ref{fig:flowfiled_laminar} shows the converged flow field contour obtained with PFD.

Subsequently, we also compare the convergence behavior of the lift coefficient, shown in Fig.\ref{fig:Cl_laminar}. The y-axis range is narrowed to observe the convergence details. Applying the PFD method every $1000$ iterations significantly reduces the oscillation amplitude of the lift coefficient. After four applications of the PFD method (that is, by $4000$ iterations), the lift coefficient exhibits virtually no oscillation and can be considered converged. In contrast, the lift coefficient computed by the original solver still exhibits noticeable oscillations at $10,000$ iterations and has not yet converged.

The convergence plots of both the residual and the lift coefficient demonstrate that the PFD method, even for this laminar case, can achieve the same converged state while substantially reducing the number of iterations, thereby providing acceleration. This further demonstrates the versatility of the PFD method.

\subsection{Case 3: Transonic turbulent flow over aircraft}

In the field of aeronautical engineering, full-aircraft CFD simulations are indispensable yet computationally intensive during aerodynamic shape design. Reducing the number of iterations required for full-aircraft CFD simulations would significantly accelerate the development cycle and reduce costs for new aircraft configurations. Therefore, to demonstrate the capability and practical engineering value of our method, this test case presents the CFD acceleration results for a three-dimensional transonic passenger aircraft benchmark model.

\begin{figure}[h]
    \centering

    \begin{subfigure}[b]{0.45\textwidth}  
        \centering
        \includegraphics[width=\textwidth]{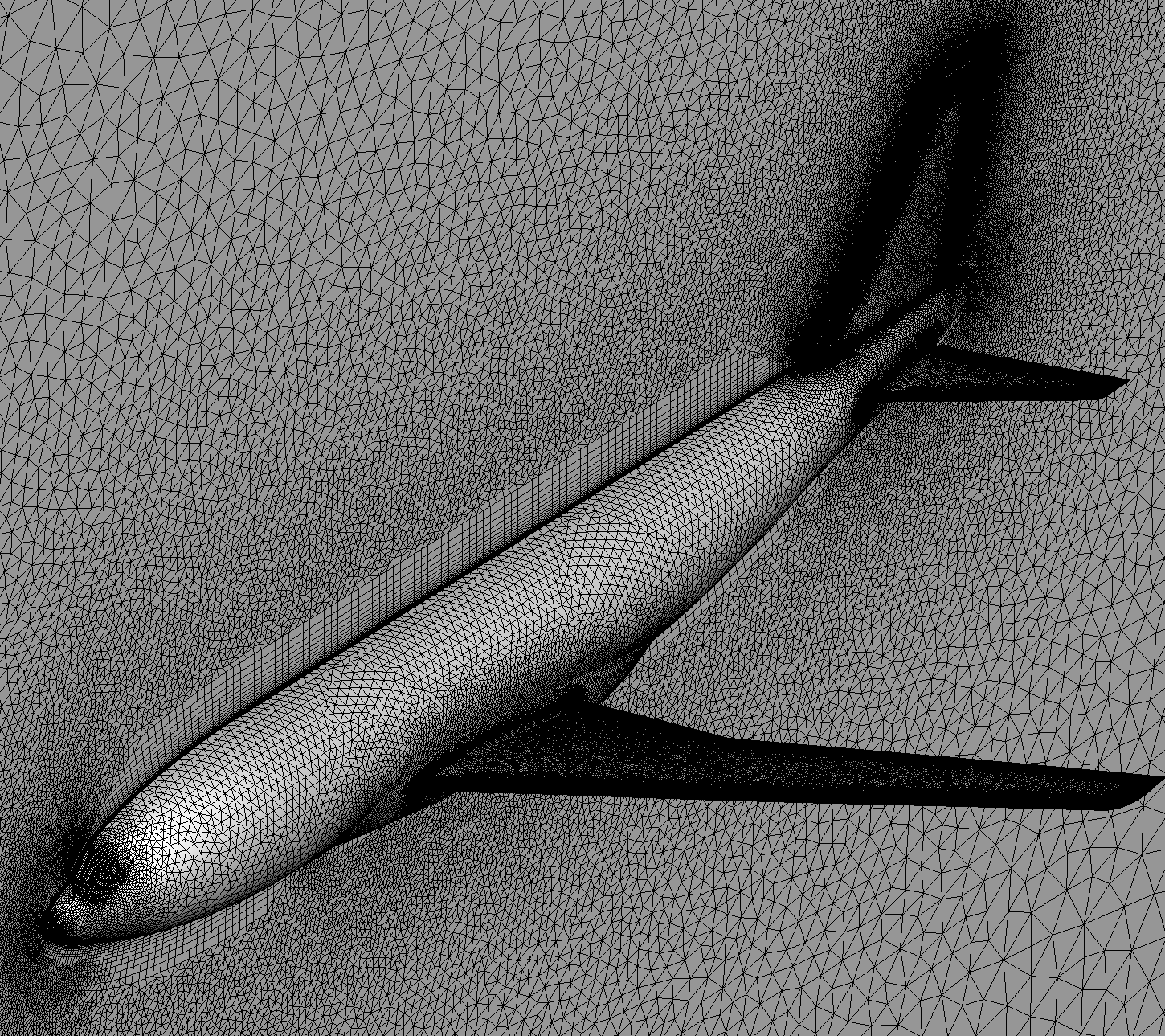}  
        \caption{Computational grid}
        \label{fig:mesh_aircraft}
    \end{subfigure}
    \hfill  
    \begin{subfigure}[b]{0.45\textwidth}
        \centering
        \includegraphics[width=\textwidth]{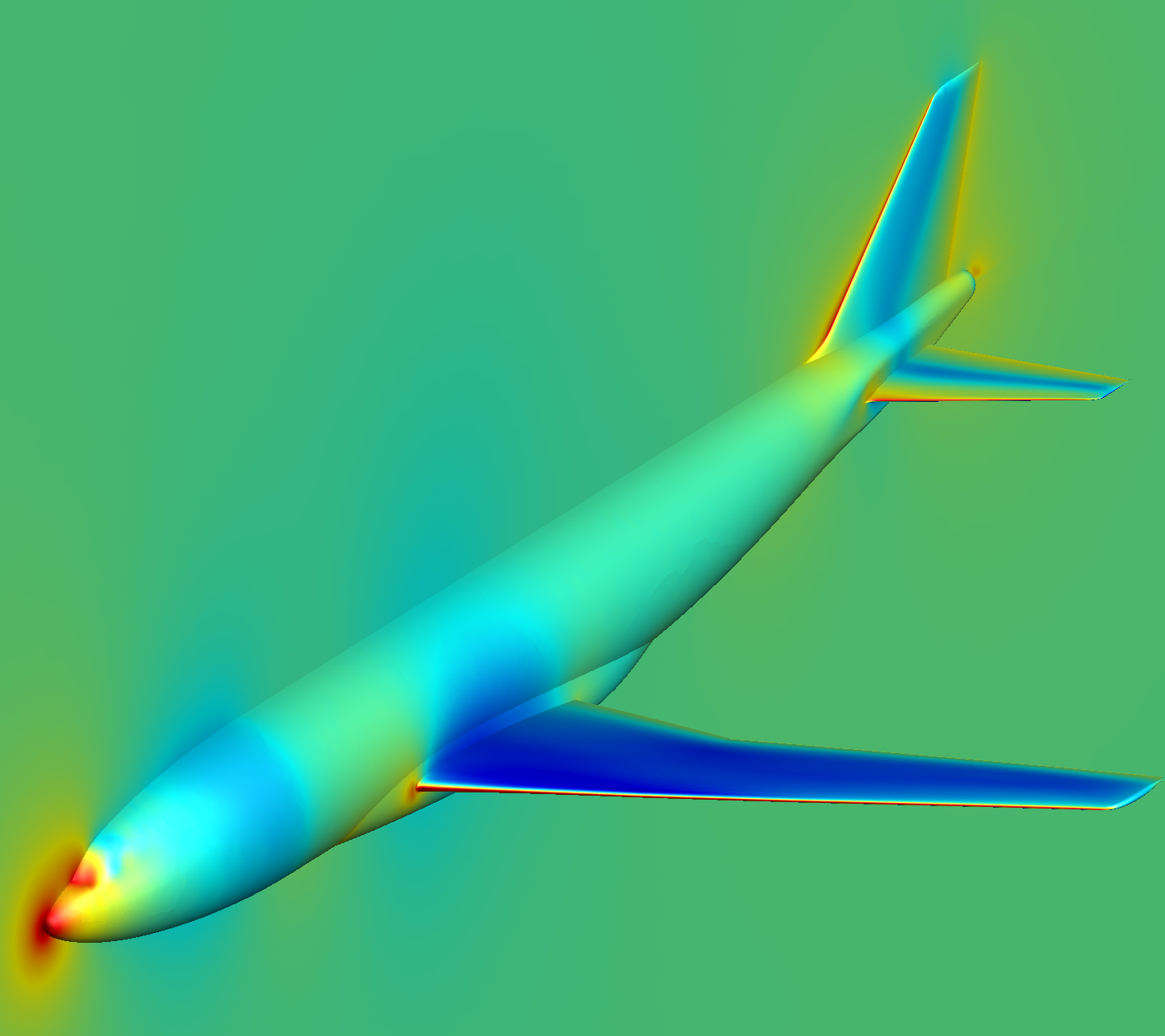}  
        \caption{Pressure distribution}
        \label{fig:flowfiled_aircraft}
    \end{subfigure}

    \caption{Computational grid and pressure distribution of transonic turbulent flow over aircraft}
    \label{fig:mesh_pressure_aircraft}
\end{figure}

Unlike the previous three cases, this case was tested using the open-source software PHengLEI. The pseudotime integration scheme is LU-SGS with a CFL number of 100, and the Spalart-Allmaras (S-A) turbulence model is employed. The freestream conditions are a Mach number of $Ma=0.78$ and a Reynolds number of $Re=1.703665 \times 10^7$. The aircraft angle of attack is $\alpha=0^\circ$ with a sideslip angle of $\beta=0^\circ$. The computational grid, shown in Fig.\ref{fig:mesh_aircraft}, consists of 21,005,478 cells and is unstructured. It encompasses the fuselage, wing, horizontal tail, and vertical tail, while the engine nacelles are excluded due to internal flow considerations. These flow conditions correspond to transonic cruise flight—the primary operational state for passenger aircraft and a representative flow regime.

\begin{figure}[h]
    \centering

    \begin{subfigure}[b]{0.48\textwidth}  
        \centering
        \includegraphics[width=\textwidth]{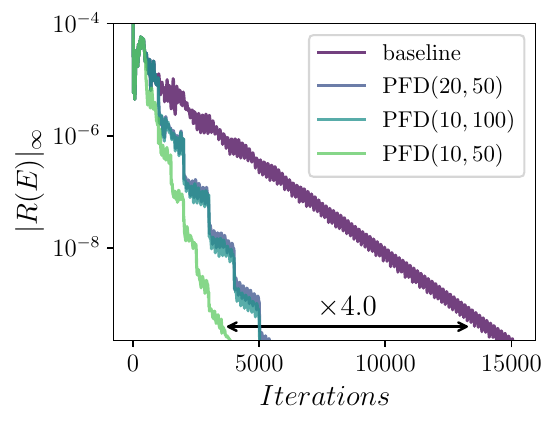}  
        \caption{Convergence histories of residual}
        \label{fig:resi_aircraft}
    \end{subfigure}
    \hfill  
    \begin{subfigure}[b]{0.5\textwidth}
        \centering
        \includegraphics[width=\textwidth]{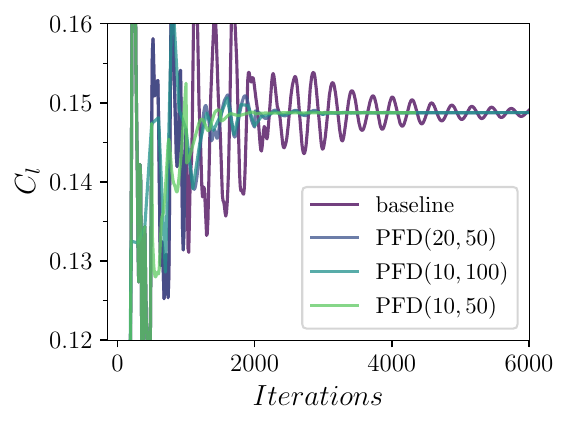}  
        \caption{Convergence histories of $C_l$}
        \label{fig:Cl_aircraft}
    \end{subfigure}

    \caption{Comparison of convergence histories of baseline method and PFD-accelerated methods for the case of transonic turbulent flow over aircraft.}
    \label{fig:convergence_histories_aircraft}
\end{figure}

The original CFD solver reached a state where the residual ceased decreasing after 15,000 iterations, deemed the converged state with a residual of $2.2 \times 10^{-13}$. This residual value serves as the convergence criterion for evaluating the acceleration capability of the PFD method. We compare the residuals of three different PFD configurations against the original baseline, as shown in Fig.\ref{fig:resi_aircraft}. From the residual convergence curves, the original solver requires 15,000 pseudo-time iterations to converge. All PFD configurations substantially reduce the required iterations. Specifically, the PFD(10,50) configuration achieves convergence in only 4,000 iterations—a reduction of 73.33\%—yielding an acceleration ratio of 3.75.

Similarly, examining the lift coefficient convergence curves (Fig.\ref{fig:Cl_aircraft}) reveals significant oscillations persist in the original solver at 7,000 iterations. In contrast, both the PFD(20,50) and PFD(10,100) configurations exhibit negligible lift coefficient oscillations beyond 4,000 iterations. The PFD(10,50) configuration achieves lift coefficient convergence in just 3,000 iterations.

Using the practically relevant engineering problem of a 3D transonic passenger aircraft benchmark model in cruise conditions, this case further validates the applicability of the PFD method under complex conditions: three-dimensional geometry, transonic flow, and high Reynolds number. This demonstrates the versatility of the PFD method across diverse CFD problems. Combined with its inherent suitability for parallelization, these results indicate the feasibility of applying PFD to practical engineering problems in the future.

\subsection{Estimate of the ratio of accelerating}

\label{sec:acc_rate_scale}
In Section \ref{sec:effects_on_spatio-temporal_system}, we define the applicable scope and acceleration principle of the PFD method. This section quantitatively evaluates its acceleration performance. The speedup ratio is defined as the number of iterations required by the baseline method divided by the number of iterations required after incorporating PFD, under identical residual error conditions.

Assuming the convergence rate of the CFD solver is governed by the term with the minimum decay rate $\sigma$ in Eq.\eqref{eq:Er}, and equating the convergence rate to $\sigma$, a single application of the PLTI method reduces both the residual and error by $-\log_{10}\Delta$ orders of magnitude. Based on this assumption and the definition of speedup ratio within the context of linear convergence, the theoretical speedup ratio is derived as:
	\begin{equation}
		Speedup\_Ratio = 1-\frac{log_{10}{\Delta}}{\sigma t_0}
		\label{eq:accratio}
	\end{equation}

The typical period for error wave propagation is approximately $T = 200-300$ iterations. Using the average value $T = 250$, $\omega$ is computed as $\omega = {2\pi}/{T} = 0.0251$. Table \ref{tab:acc_ratio_contrast} compares the theoretical speedup ratios calculated using Eq.\eqref{eq:accratio} against actual speedup ratios observed in the first three CFD test cases presented in the following chapter:

	\begin{table}
		\begin{center}
			\caption{Comparison of Speedup Ratios ($\omega = 0.0251$)}
			\label{tab:acc_ratio_contrast}
			\begin{tabular}{l|c|r} 
				\textbf{} & \textbf{Eq. \eqref{eq:accratio}} & \textbf{Actual}\\
				\hline
				Case 1 ($t_0=400,\text{CFL}=2$) & 1.76 & 2.20\\
				Case 1 ($t_0=500,\text{CFL}=2$) & 1.77 & 2.20\\
				Case 2 ($t_0=1000,\text{CFL}=2$) & 2.78 & 3.08\\
				Case 2 ($t_0=500,\text{CFL}=100$) & 1.56 & 2.0\\
                Case 2 ($t_0=500,\text{CFL}=1000$) & 1.56 & 2.0\\
				Case 3 ($t_0=500,\text{CFL}=100$) & 3.43 & 3.75\\
				Case 3 ($t_0=1000,\text{CFL}=100$) & 2.94 & 3.00\\
				
			\end{tabular}
		\end{center}
		
	\end{table}

\subsection{Case 4: Solving systems of linear equations}

\begin{figure}[h]
    \centering

    \begin{subfigure}[b]{0.5\textwidth}  
        \centering
        \includegraphics[width=\textwidth]{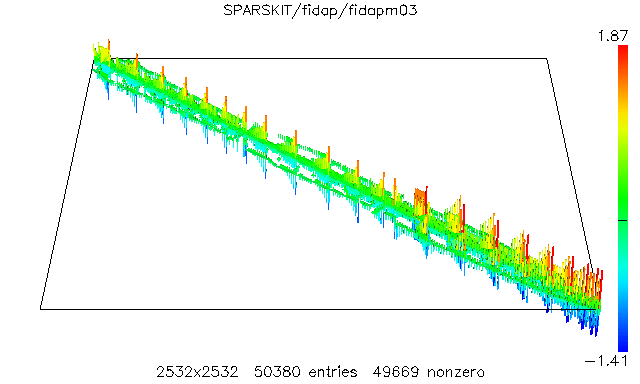}  
        \caption{City plot of matrix FIDAPM 03}
        \label{fig:cityplot}
    \end{subfigure}
    \hfill  
    \begin{subfigure}[b]{0.48\textwidth}
        \centering
        \includegraphics[width=\textwidth]{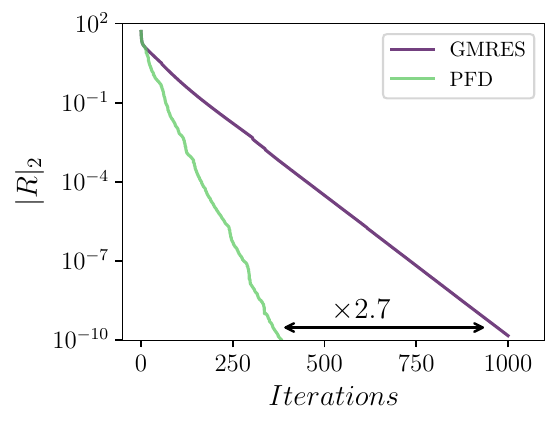}  
        \caption{Comparison of convergence histories.}
        \label{fig:resi_gmres}
    \end{subfigure}

    \caption{(a) City plot of matrix FIDAPM 03 and (b) comparison of convergence histories.}
    \label{fig:matrix}
\end{figure}

In numerical computation, complex computational problems are often ultimately reduced to solving systems of linear equations $A\mathbf{x} = \mathbf{b}$. Solving $A\mathbf{x} = \mathbf{b}$ is a fundamental task encountered in numerical computation across various engineering disciplines. Consequently, accelerating the solution of linear systems has become a common challenge in scientific computing.

For solving linear systems corresponding to large, sparse, non-symmetric matrices $A$, the Generalized Minimal Residual (GMRES) method is currently the most prevalent and versatile approach.GMRES approximates the solution using a Krylov subspace and minimizes the norm of the residual vector iteratively, ultimately achieving convergence. However, GMRES requires storing the Krylov subspace basis vectors. The computational complexity of the Arnoldi process within GMRES is $O(n^2)$, where $n$ is the dimension of the Krylov subspace. Due to practical limitations in memory and central processing unit (CPU) resources, storing an unlimited number of Krylov vectors is infeasible. Therefore, in practical applications, especially for solving large systems, the GMRES method must be restarted after a certain number of iterations. A critical factor for the success of the restarted iteration process is obtaining an initial guess with a smaller residual and improved stability upon restart.

In this test case, the FIDAPM 03 matrix from the Matrix Market collection is selected as matrix $A$. This matrix has dimensions $2532 \times 2532$, is real-valued and non-symmetric, and is not diagonally dominant. Its structure plot and city plot are shown in Fig.\ref{fig:cityplot}. The standard GMRES method was restarted every 100 iterations. The PFD method collects historical snapshots generated at each iteration step and utilizes them during the GMRES restart, corresponding to the general format PFD-1-100. Fig.\ref{fig:resi_gmres} compares the residual convergence history of the original GMRES method with GMRES augmented by PFD. The original GMRES method required 100,000 iterations for the 2-norm of the residual $\mathbf{r} = A\mathbf{x} - \mathbf{b}$ to converge to $10^{-10}$. In contrast, GMRES utilizing PFD during restarts achieved this convergence criterion in only 37,000 iterations, saving 63\% of the iterations and yielding an acceleration ratio of 2.70. This demonstrates that PFD provides an initial guess with a smaller residual and enhanced stability at restart, significantly simplifying the subsequent iteration process and thereby achieving acceleration.

This case study provides a detailed investigation into the acceleration effect of PFD when applied to the mainstream GMRES method for solving linear systems $A\mathbf{x} = \mathbf{b}$. The results indicate that the PFD method is not only effective for accelerating CFD solution processes but also holds promise for application to the fundamental problem of solving $A\mathbf{x} = \mathbf{b}$, a task common to many scientific computing domains. This further demonstrates the broader applicability potential of PFD across different computational physics problems.

\subsection{Case 5: Reduction of force coefficient errors}

\begin{figure}[h]
    \centering

    \begin{subfigure}[b]{0.49\textwidth}  
        \centering
        \includegraphics[width=\textwidth]{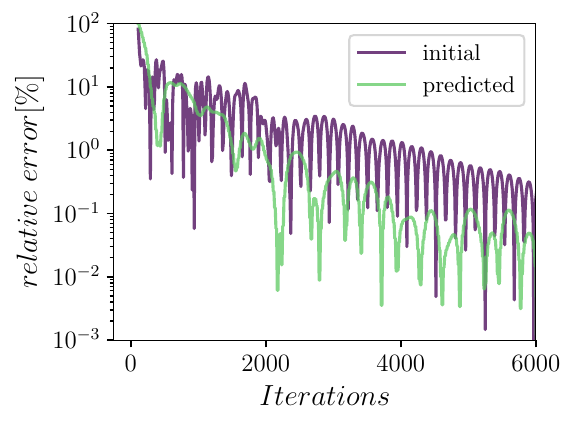}  
        \caption{$C_l$}
        \label{fig:C_l}
    \end{subfigure}
    \hfill  
    \begin{subfigure}[b]{0.49\textwidth}
        \centering
        \includegraphics[width=\textwidth]{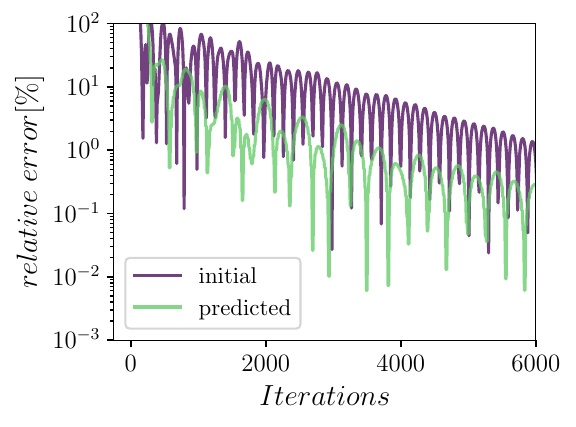}  
        \caption{$C_d$}
        \label{fig:C_d}
    \end{subfigure}

    \caption{Convergence histories of (a)$C_l$ and (b)$C_d$.}
    \label{fig:C_landC_d}
\end{figure}

The preceding section demonstrates that utilizing the truncated form of Eq.\ref{eq:formula1} to predict function limits can be employed to make iterative parameter values approach convergence values more closely. Naturally, it can also be directly applied to convergence curves of flow field results, such as force coefficients, to reduce their errors. Here, we express Eq.\ref{eq:formula1} as:
	\begin{equation}
		f(+\infty) = f(0^+) - \frac{\int_{0^+}^{+\infty} \frac{f(x)-f(\frac{1}{2}x)}{x} \, dx}{\ln \frac{1}{2}}
		\label{eq:ForcePrediction}
	\end{equation}
and write it in the following truncated form:
	\begin{equation}
		g(x) = f(0^+) - \frac{\int_{0^+}^{x} \frac{f(x)-f(\frac{1}{2}x)}{x} \, dx}{\ln \frac{1}{2}}
		\label{eq:ForcePredictionTruncated}
	\end{equation}

This test case demonstrates the error-reduction effect of applying the PFD method to the original convergence curves of the lift and drag coefficients for the aircraft (Case 3), as shown in Fig.\ref{fig:C_landC_d} (compared to Fig.\ref{fig:Cl_aircraft}).

In the figure, the horizontal axis represents the pseudo-time iteration step, and the vertical axis represents the percentage error. The error is defined as the relative difference compared to the force coefficient value at full convergence. The original force coefficient convergence curve (Initial) stores one data point every 10 iteration steps. The PFD curve is generated by processing the original data according to Eq.\ref{eq:g} at different values of x (i.e., at different iteration steps).

Fig.\ref{fig:C_l} compares the lift coefficient errors, and Fig.\ref{fig:C_d} compares the drag coefficient errors. It can be seen that after processing with Equation \ref{eq:g}, the PFD error curves for both coefficients are reduced by approximately an order of magnitude compared to the initial error curves. Furthermore, the PFD curves achieve error levels comparable to those reached by the Initial curve 3000-4000 iterations later. This implies that, using only the original convergence data and at virtually no additional computational cost, we can obtain force coefficients with errors equivalent to those obtained after 3000-4000 additional iterations. This demonstrates that the PFD method can function as a post-processing scheme to directly reduce errors in flow field results, thereby effectively reducing the required number of computational iterations.

\section{Conclusions}

This paper first proposes a novel data-driven approach for accelerating convergence: point-wise prediction of the limit values for individual parameters. This acceleration methodology distinguishes itself from previous approaches that apply uniform processing to the entire field. It inherently guarantees consistency between serial and parallel implementations due to its fundamental mechanism. Since predicting the limit value of a single parameter is involved, and the temporal evolution of a single parameter constitutes a univariate function, establishing an integral formula linking historical information to the limit value becomes a natural process.

Subsequently, we discretized the integral formula and found it could be simplified into the most elementary purely algebraic expression. We discussed the applicability of this integral formula to univariate functions. Through extensive derivation, we demonstrated its capability to significantly reduce the error of fundamental oscillatory error wavefunctions (e.g., $e^{-\beta t}\cos(ax+bt)$) when spatial dimensions are incorporated, elucidating the acceleration mechanism of the method. This ultimately yielded a quantitative expression for predicting the acceleration ratio. Comparisons across several test cases confirmed that this expression exhibits excellent agreement with the actual measured acceleration ratios.

Finally, we presented numerous test cases demonstrating the effectiveness of the PFD method, achieving acceleration ratios between 2.5 and 4. The acceleration ratio of up to 4 obtained for the three-dimensional transonic passenger aircraft simulation particularly underscores the high efficiency of our approach. Combined with the inherent parallelizability afforded by its point-wise operation and the simple algebraic form of PFD, the method is exceptionally well-suited for large-scale, parallel, complex engineering simulations.

\section*{Acknowledgments}
Xukun Wang acknowledge the financial support of the China Scholarship Council(CSC).


\bibliographystyle{elsarticle-num-names} 
\bibliography{references}



\end{document}